\documentstyle[aps,epsf]{revtex}
\begin{document}

\title{Jets and produced particles in $pp$ collisions
       from SPS to RHIC energies \\ for nuclear applications}

\author{G.G. Barnaf\"oldi$^{1}$, G. Fai$^{2}$, P. L\'evai$^{1,2}$,
 G. Papp$^{2,3}$, Y. Zhang$^{2}$}
\address{
$^{1}$ KFKI Research Institute for Particle and Nuclear Physics, 
P. O. Box 49, Budapest 1525, Hungary \\
$^{2}$ Physics Department, Kent State University, Kent OH 44242, USA \\
$^{3}$ HAS Research Group for Theoretical Physics, E\"otv\"os University,
P\'azm\'any P. 1/A, Budapest 1117, Hungary
}

\date{\today}
\maketitle

\vspace*{-4.5cm}
\begin{flushright}
{KSUCNR-103-00}
\end{flushright}
\vspace*{3.5cm}
\begin{abstract}
Higher-order pQCD corrections play an important role in the
reproduction of data at high transverse momenta
in the energy range 20 GeV $ \leq \sqrt{s} \leq 200$ GeV.
Recent calculations of photon and pion production in $pp$ collisions
yield detailed information on the next-to-leading order
contributions. However, the application of these results in 
proton-nucleus and nucleus-nucleus collisions is not straightforward.
The study of nuclear effects requires
a simplified understanding of the output of these 
computations. Here we summarize our analysis of recent
calculations, aimed at handling the NLO results
by introducing process and energy-dependent $K$ factors. \\

\end{abstract}


The main motivation of ultrarelativistic heavy-ion
collision experiments is their role in testing quantum chromodynamics (QCD)
and, in particular, their potential to display
the predicted phase transition between hadron matter and
the quark-gluon plasma (QGP)\cite{harris}. 
At higher transverse momenta ($p_T \gtrsim 3$~GeV),
QGP scenarios should be judged against the background of
basic particle yields from primary nucleon-nucleon collisions,
which can be described by perturbative QCD (pQCD)\cite{field89}. 
Thus, to verify the formation
of the QGP state\cite{qm99}, we need precise knowledge about the
`partonic background' provided by nucleon-nucleon collisions.

The pQCD-based description of particle production in proton-proton 
($pp$) collisions is a challenging task. Using pQCD at relatively low energy
further complicates the situation. Recent studies with 
higher order contributions\cite{Aur99f,Aur99p,kidonakis99,florian99} 
investigated the role of different QCD scales in detail.
However, the application of these results
as a pQCD reference is not an easy job for several reasons.
First, a recent NLO pion calculation has difficulty in consistently
reproducing the pion data\cite{Aur99p}. 
Second, although still of relatively low order,
these calculations are rather long and complicated. 
It would be beneficial to find a fast phenomenological
shortcut with appropriate precision.
Finally, a heavy ion 
collision is not a mere superposition of binary nucleon-nucleon
collisions, but the incoming nucleons and the produced partons
are subject to multiple scattering and to other medium effects 
(e.g. higher-twist\cite{guo96}, shadowing\cite{eskola96})
due to the presence of further nucleons 
or partons. The phenomenological description should have the 
flexibility of being able to include many-body effects.

The first and widely-used step in creating such an effective pQCD 
description is the introduction of the so-called ``$K$ factor'', which,
roughly speaking, 
renormalizes the leading term of the pQCD cross section and 
accounts for higher-order corrections in different 
processes\cite{EskWang95,wong98}. The higher-order 
corrections are known to be large and 
process dependent\cite{Soper1,Soper2,Kdrell}. Note that 
several different $K$ factors exist in the literature. For example, 
from an experimental point of view, it is worthwhile to compare data 
to a given order of a pQCD calculation, and this is frequently expressed 
in terms of a $K$ factor. Theoretically, the $K$ factor can be thought of 
as the ratio of a full pQCD cross section to a cross section through 
given order, or as the ratio of calculations through different orders,
or the $K$ factor can be studied in the framework of a model.
A study of the $K$ factor in the framework of the Parton Cascade 
Model (PCM) was carried out recently\cite{Bass}.

Here we insist on defining the $K$ factor {\it within} a given order
of pQCD (specifically NLO) as the factor correcting
the Born-approximation result in NLO to the full NLO result for the
particular process. In any pQCD calculation the partonic
cross section is convoluted with parton distribution functions, which 
describe some of the perturbatively incalculable soft physics. 
The NLO parton distribution functions are fitted to the data
in the framework of an NLO calculation\cite{MRST98}. For consistency, it
is important that parton distribution functions of the 
order of the cross-section calculation be used. The $K$ factor 
can be introduced as:
\begin{equation}
\frac{d\sigma^{NLO}}{d^3p} = \frac{d\sigma^{Born}}{d^3p} + 
\frac{d\sigma^{corr}}{d^3p} = K(s,p_T) \frac{d\sigma^{Born}}{d^3p} \ .
\label{kdef}
\end{equation}
In general, the $K$ factor is energy and momentum dependent. 
However, in many cases it is approximated as a constant in a 
certain momentum region for a given energy.
The $K$ factor is traditionally thought to have values 
in the range of 1.5 -- 2.5. 

In this paper we summarize our recent results on the value and the behavior
of the $K$ factor in the energy region 20 GeV $ \leq \sqrt{s} 
\leq 200$ GeV. Within this interval,
we concentrate on the lower energies, where proton-nucleus data
are abundant. First we display our results on the ``jet level''.
We then investigate the modifications due to jet fragmentation
and especially to pion production. We close with an
overview of recent results on photon production. A forthcoming 
publication\cite{pzfbl00} will include comparisons to experimental data
using $K$ factors from the present paper.

The computation of the partonic cross section 
through NLO is tedious. The calculations
by the Ellis-Kunszt-Soper (EKS) group\cite{Soper1,Soper2} and by the
Aversa-Chiappetta-Greco-Guillet group\cite{Aversa}
are based on the matrix elements published in Ref.~\cite{Ellis}.
We use EKS's public FORTRAN-code\cite{EKScode} to calculate the
Born and correction contributions in our jet-level analysis.
For the parton distribution functions 
we use the MRST (central gluon) set\cite{MRST98}, which incorporates a
large amount of experimental information evaluated at the NLO level.

The choice of the factorization/renormalization scale $Q$ 
influences both the LO and the NLO jet cross sections.
The typical value of this parameter varies around
$p_T/3 \lesssim Q \lesssim 2 p_T$ in standard calculations.
Here we take $Q = p_T/2$. (Naturally, $p_T$ refers to the
transverse momentum of the jet in jet-level expressions, while it
denotes the transverse momentum of the final hadron when observable
hadron production is calculated.)  It should be kept in mind that several 
complications arise in NLO which do not plague the LO calculation.
Among these are the dependence of the NLO results on the jet-cone
angle, $R = \sqrt{(\Delta\eta)^2 + (\Delta\Phi)^2}$ (with pseudorapidity
$\eta$ and azimuth $\Phi$)
and the jet-separation parameter $R_{sep}$. Following EKS\cite{EKScode},
the collinear singularity, which also appears first in NLO, is treated 
here by choosing
$Q_{collinear} = p_T$. We are particularly interested in the 
dependence on $R$, and keep $R_{sep}=2 R$ in our calculations.
We display two results. In Fig. 1
we use the `optimal size' $R=0.7$ for the jet cone, where the scale
dependence on the factorization scale $Q$ is minimal
\cite{EKS92,Huston}.
Fig. 2 displays results with a larger cone size, $R=1.0$, where
the bremsstrahlung contributions are enhanced, causing
an increase of the $K$ factor. 
In both figures we show $K$ as a function of
$p_T$ for energies 20~GeV $\leq \sqrt{s} \leq$ 200 GeV. We present 
calculated results for the $p_T$ region where the accuracy of the 
code is better than 5\%. The solid
lines represent a fit to the calculated data points at the particular energy
(linear fit in Fig.1, quadratic in Fig. 2).
It can be seen in both figures that the magnitude and the 
transverse-momentum dependence 
of the $K$ factor decreases with increasing energy. At $\sqrt{s} =$ 200 GeV,
the value varies less than 5\% from $p_T = 3$~GeV to $p_T = 10$ GeV.
In contrast, at $\sqrt{s} =$ 20 GeV, the variation is on the order of
20\% from $p_T = 2.5$ GeV to $p_T = 4$ GeV. This is not unexpected,
since higher-order terms should become less important with increasing 
energy. Comparing Fig.-s 1 and 2, we see 
that the $K$ factor is larger if $R$ is larger and the dependence of 
$K$ on $p_T$ is more linear with the smaller value of $R$. This is due to
the increased contribution of higher values of $p_T$  with increasing 
cone size. For practical purposes, we 
parameterize the energy and momentum dependence of the
$K$ factor for $R=0.7$ as
\begin{equation} 
K_{jet}(s,p_T) = 
1. + \frac{65.}{\sqrt{s}+160.} + 
\frac{2.}{\sqrt{s}-6.}\,\,p_T   \ \,\,  .
\label{kjet}
\end{equation}
For $R=1.0$ we find
\begin{equation}
K_{jet}(s,p_T) = 
1.6 + \frac{20.}{\sqrt{s}} - \frac{24.}{(\sqrt{s}-10.)^2}\,\,p_T
			   + \frac{ 6.}{(\sqrt{s}-10.)^2}\,\,p_T^2  \,\,\, .					
\label{kjet2}
\end{equation}
These parameterizations are also shown on Fig.-s 1 and 2 (dashed lines).

So far, we only discussed the $K$ factor at the jet level. The above
parameterization characterizes
an average enhancement of jet production through NLO.
To calculate {\it observable hadron
production} via parton fragmentation in an 
NLO treatment, one may substitute the above 
$K_{jet}(s,p_T)$ into the pQCD calculation
leading to an approximate NLO-result:
\begin{eqnarray}
\label{hadX}
  E_{h}\frac{d\sigma_h^{pp}}{d^3p} &=&
        \sum_{abcd} \int\!dx_a dx_b\ f^{NLO}_{a/p}(x_a,Q^2)\
        f^{NLO}_{b/p}(x_b,Q^2)\ \nonumber \\
&& \left[ K_{jet}(s, p_{T,c}) 
   \frac{d\sigma^{Born}}{d{\hat t}}(ab \to cd)\, \right]\  
   \frac{D_{h/c}^{NLO}(z_c, Q'^2)}{\pi z_c} \ \ \  ,
\end{eqnarray}
where $f^{NLO}_{a/p}(x_a,Q^2)$ stands for the parton distribution function
and $D_{h/c}^{NLO}(z_c, Q'^2)$ denotes the fragmentation function 
(both in NLO). The label $c$ represents the fragmenting parton.
The fragmentation scale is taken to be $Q'=p_T/2$, while the factorization
scale is kept at $Q=p_{T,c}/2$.
Applying this approximation and using a set of 
fragmentation functions\cite{BKK},
one can easily determine e.g. pion and kaon production through
NLO, and extract an appropriate $K$ factor for (average) {\it pions},
$K_{\pi}(s,p_T)$, defined as the ratio of the cross section 
in Eq. (\ref{hadX}) to the cross section with $K_{jet} = 1$. 
This is displayed in Fig. 3, together with a 
similarly-defined $K_{K}(s,p_T)$, in the 
energy range 20~GeV $\leq \sqrt{s} \leq $ 200 GeV. We obtained
very similar pion and kaon $K$ factors.
One advantage of using $K_{jet}$ is that a single approximate 
function can be applied in
calculating the cross sections for several similarly-produced hadrons.

The $K$ factor at the hadron level is larger than its jet-level
counterpart by 10 -- 20\%. This difference can be thought of  
as a `backshift' in $p_T$, as some of the transverse momentum of
the jet manifests itself in other particles after hadronization.
Fig. 4. illustrates these results at the lowest energies
considered, $\sqrt{s} = 24$ GeV. Extracted $K_{jet}$ (dotted lines)
and $K_{\pi}$ (solid lines) as functions of $p_T$ are displayed 
for $R=0.7$, 0.85, and 1. It can be seen that the pion
$K$ factor is approximately parallel to the corresponding jet-level 
$K$ factor, shifted toward lower values of $p_T$, as discussed above. 

Fig. 5 completes the picture. Here we present calculated  
$K$ factor for photons, based on Ref. \cite{Aur99f,Moncode},
using Eq. (\ref{kdef}), in the same energy and 
transverse momentum range. It is important to note that
this calculation includes the contribution from 
jet fragmentation\cite{Aur99f}. We see that the $K_{\gamma}$ obtained in
this manner shows a similar tendency to the $K$ factor with 
$R=0.7$ (Fig. 1).

These results suggest that   
an averaged, approximate version of an NLO calculation represented by
a $K(s,p_T)$ factor will be useful in heavy-ion collisions in this 
transverse-momentum range from SPS to RHIC energies. The above 
parameterizations (or similar ones at different $R$) can be utilized
together with the Born calculation in this region. 
As an application, and as an 
improvement on our earlier work\cite{plf00}, we intend to use
these $K$ factors in the energy range 20 GeV $\leq \sqrt{s} \leq$ 60 GeV,
at transverse momenta 3~GeV $\leq p_T \leq$ 7 GeV\cite{pzfbl00}.
At RHIC energies ($\sqrt{s} = 140 - 200$ GeV) we find an approximately 
constant $K$ factor. The numerical value with $Q = p_T/2$ and $R = 0.7$
is $K \approx 1.2$.

In conclusion, we have developed a simplified phenomenological 
description of recent NLO results for particle production in 
$pp$ collisions in terms of energy and transverse-momentum dependent 
$K$ factors. Since the complexity of proton-nucleus and 
nucleus-nucleus collisions presents its own challenges, we anticipate
that our parameterization of the $K$ factor will find useful 
applications in the analysis of forthcoming data.

\acknowledgements
We thank M. Werlen, P. Aurenche, and D. Soper for stimulating 
discussions. This work was supported in part
by U.S. DOE grant DE-FG02-86ER-40251, Hungarian 
OTKA Grant No. T032796, FKFP grant 0220/2000,
and by the US-Hungarian Joint Fund No. 652.
Partial support by the Domus Hungarica program of the Hungarian 
Academy of Sciences and by the Research Council of Kent State 
University is gratefully acknowledged.

\newpage
\begin{center}
\vspace*{18.0cm}
\includegraphics{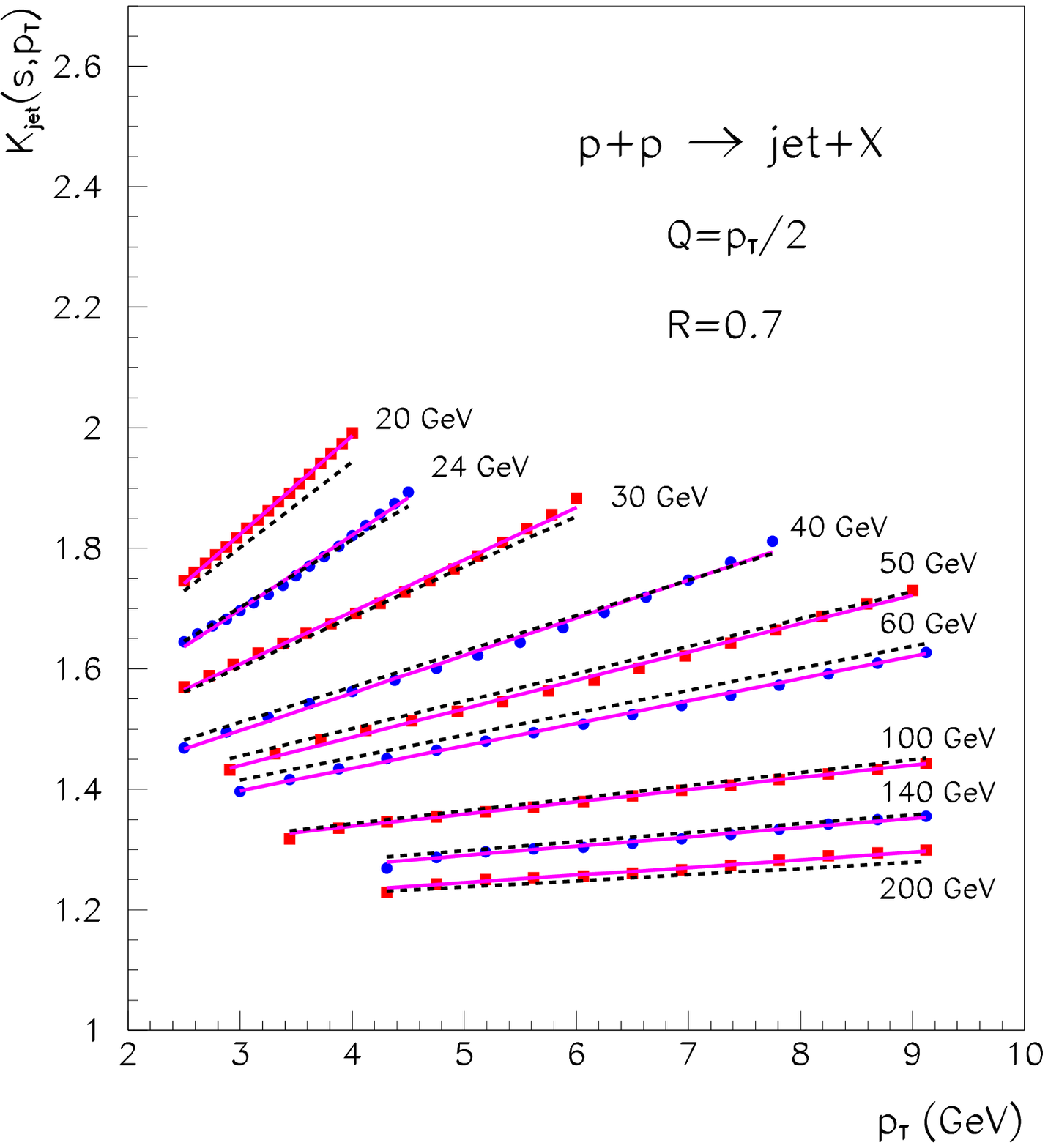}
\begin{minipage}[t]{15.cm}
{ {\bf Fig.~1.}
Energy and transverse-momentum dependent $K$ factor, $K_{jet}(s,p_T)$  
with $Q=p_{T}/2$, $R=0.7$, $R_{sep}=2R$. Dashed lines represent
Eq. (\ref{kjet}).
}
\end{minipage}
\end{center}

\begin{center}
\vspace*{18.0cm}
\includegraphics{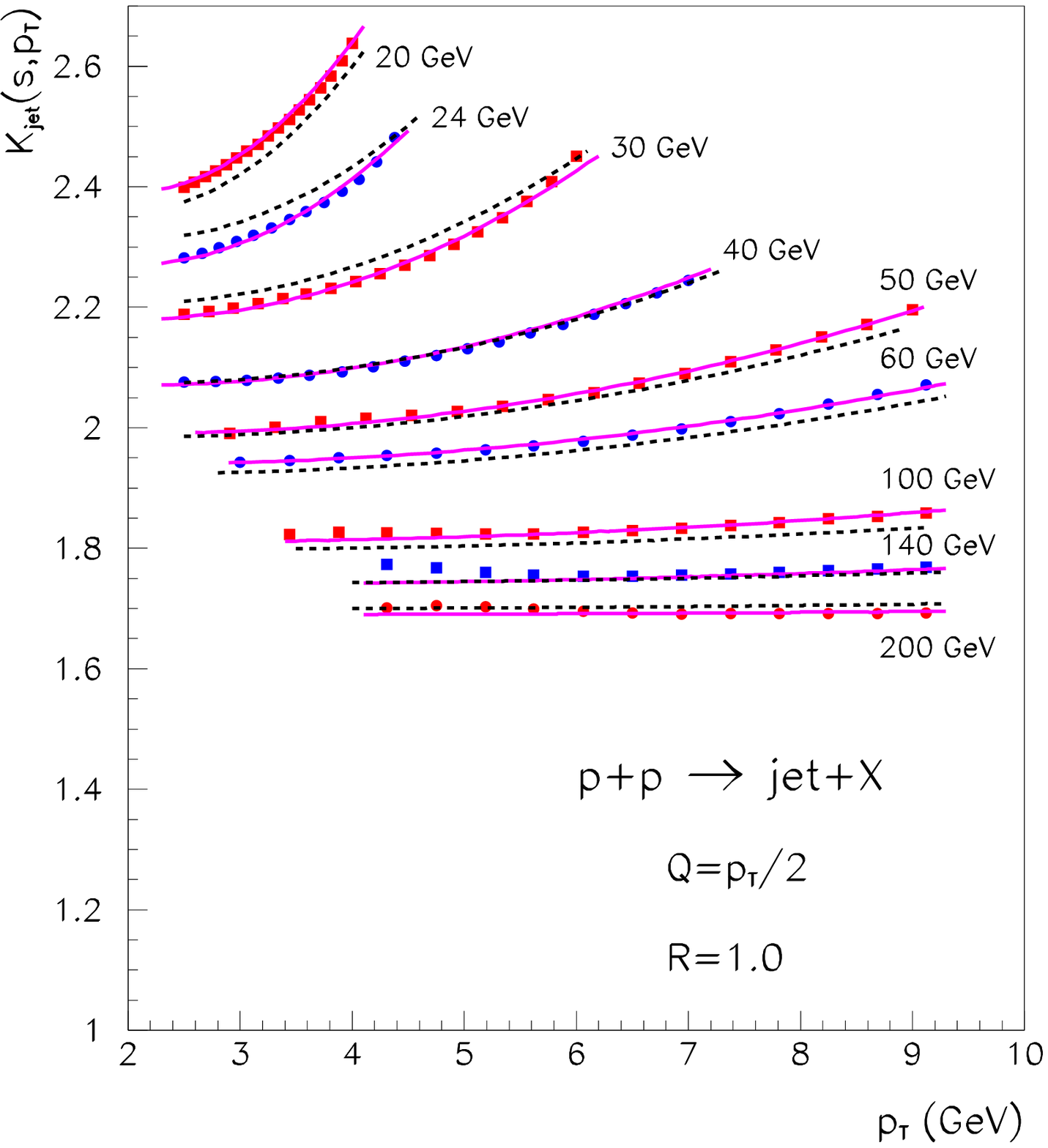}
\begin{minipage}[t]{15.cm}
{ {\bf Fig.~2.}
$K$ factor, $K_{jet}(s,p_T)$ 
with  $R=1.0$ ($Q=p_{T}/2$ and $R_{sep}=2R$). Dashed lines 
correspond to Eq. (\ref{kjet2}).
}
\end{minipage}
\end{center}

\newpage
\begin{center}
\vspace*{18.0cm}
\includegraphics{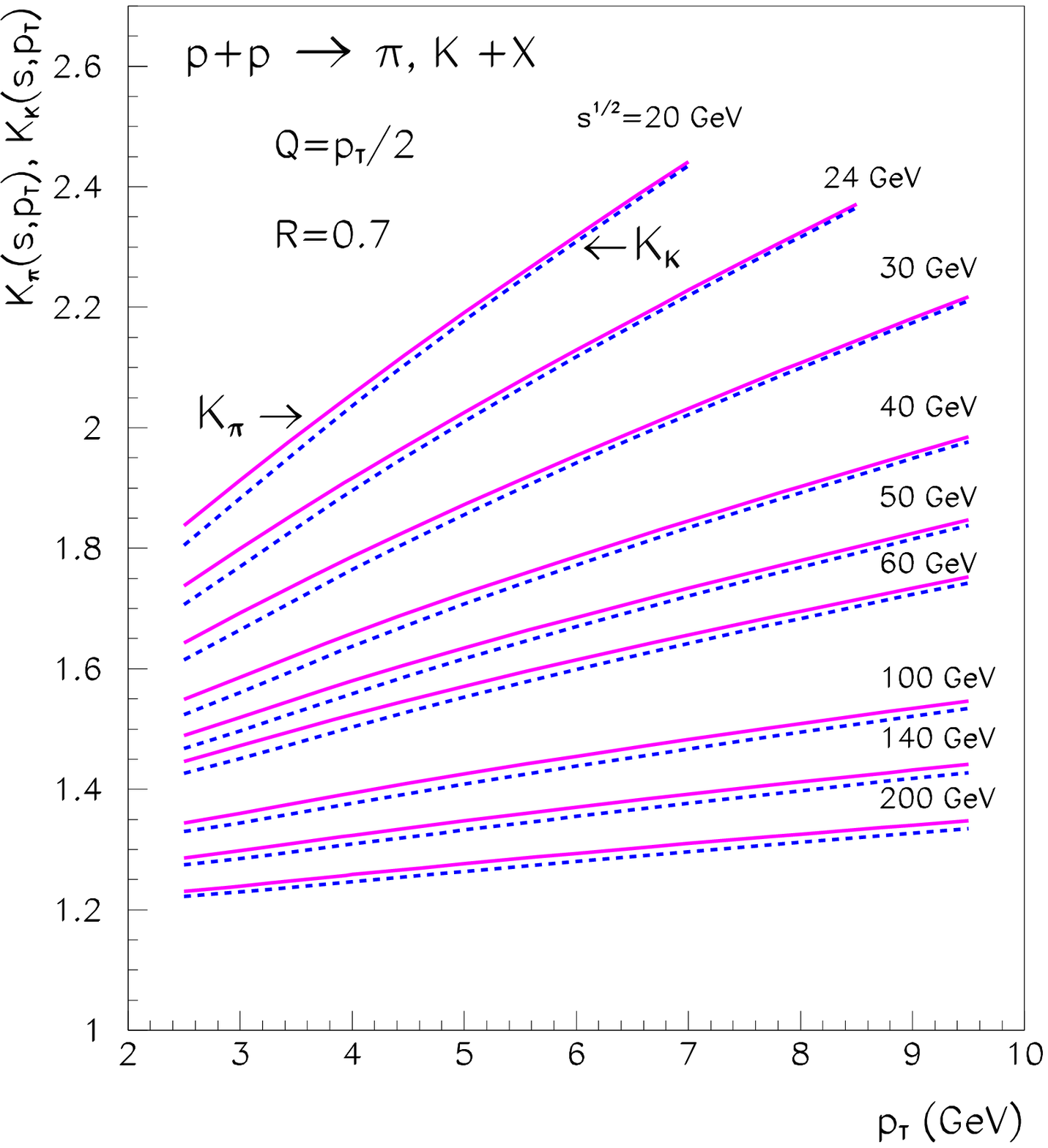}
\begin{minipage}[t]{15.cm}
{ {\bf Fig.~3.}
$K$ factor for pions, $K_{\pi}(s,p_T)$ (solid line) and 
for kaons, $K_{K}(s,p_T)$ (dashed), after hadronization
of jets calculated with $R=0.7$ ($Q=p_{T}/2$, $R_{sep}=2R$).
}
\end{minipage}
\end{center}

\begin{center}
\vspace*{18.0cm}
\includegraphics{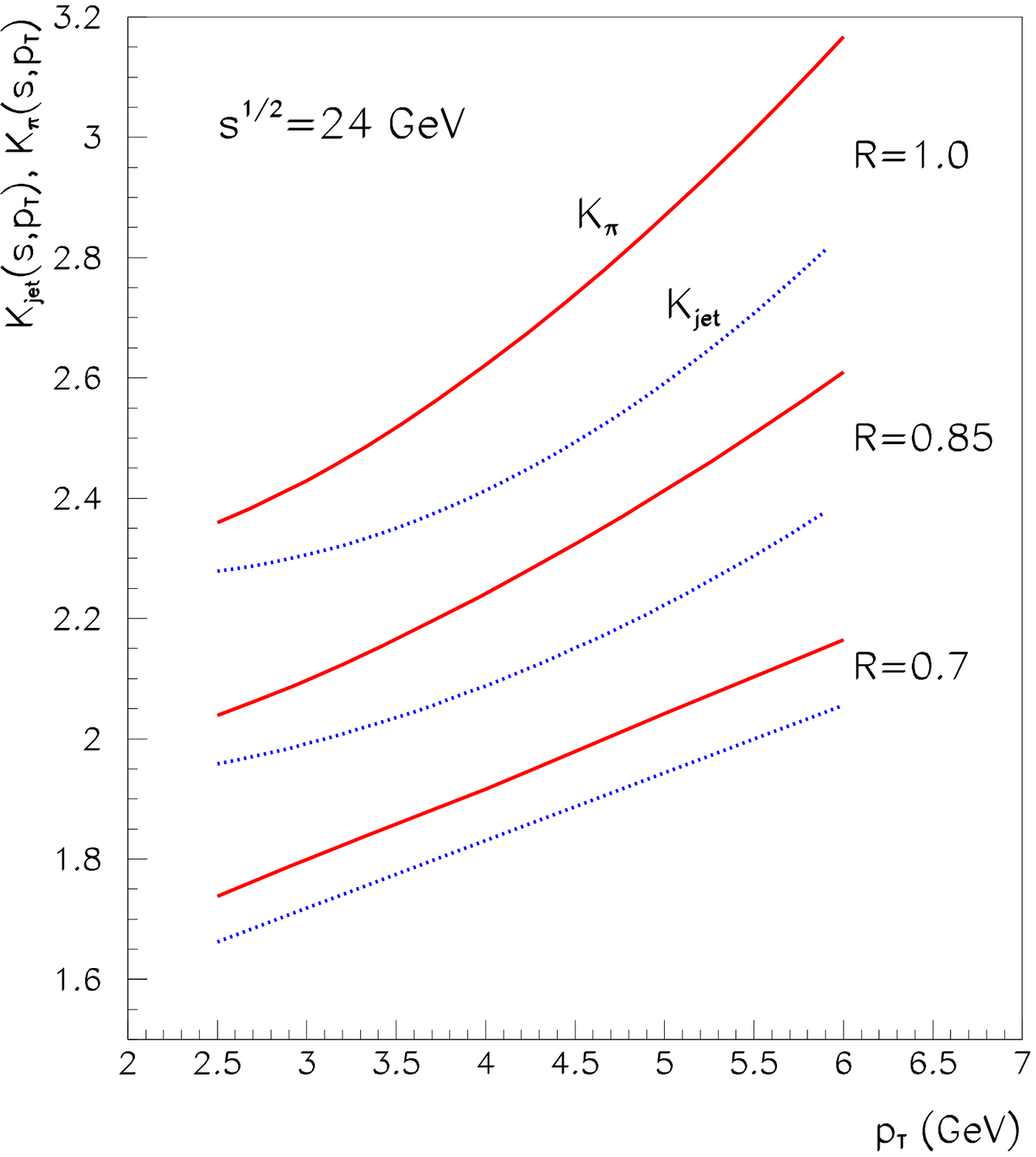}
\begin{minipage}[t]{15.cm}
{ {\bf Fig.~4.}
Comparison of $K$ factors,
$K_{jet}(s,p_T)$ (dotted lines) and $K_{\pi}(s,p_T)$ (full lines)
at $\sqrt{s}=24$ GeV, 
with $R=0.7$, $R=0.85$ and $R=1.0$ ($Q=p_{T}/2$, $R_{sep}=2R$).
}
\end{minipage}
\end{center}

\newpage
\begin{center}
\vspace*{18.0cm}
\includegraphics{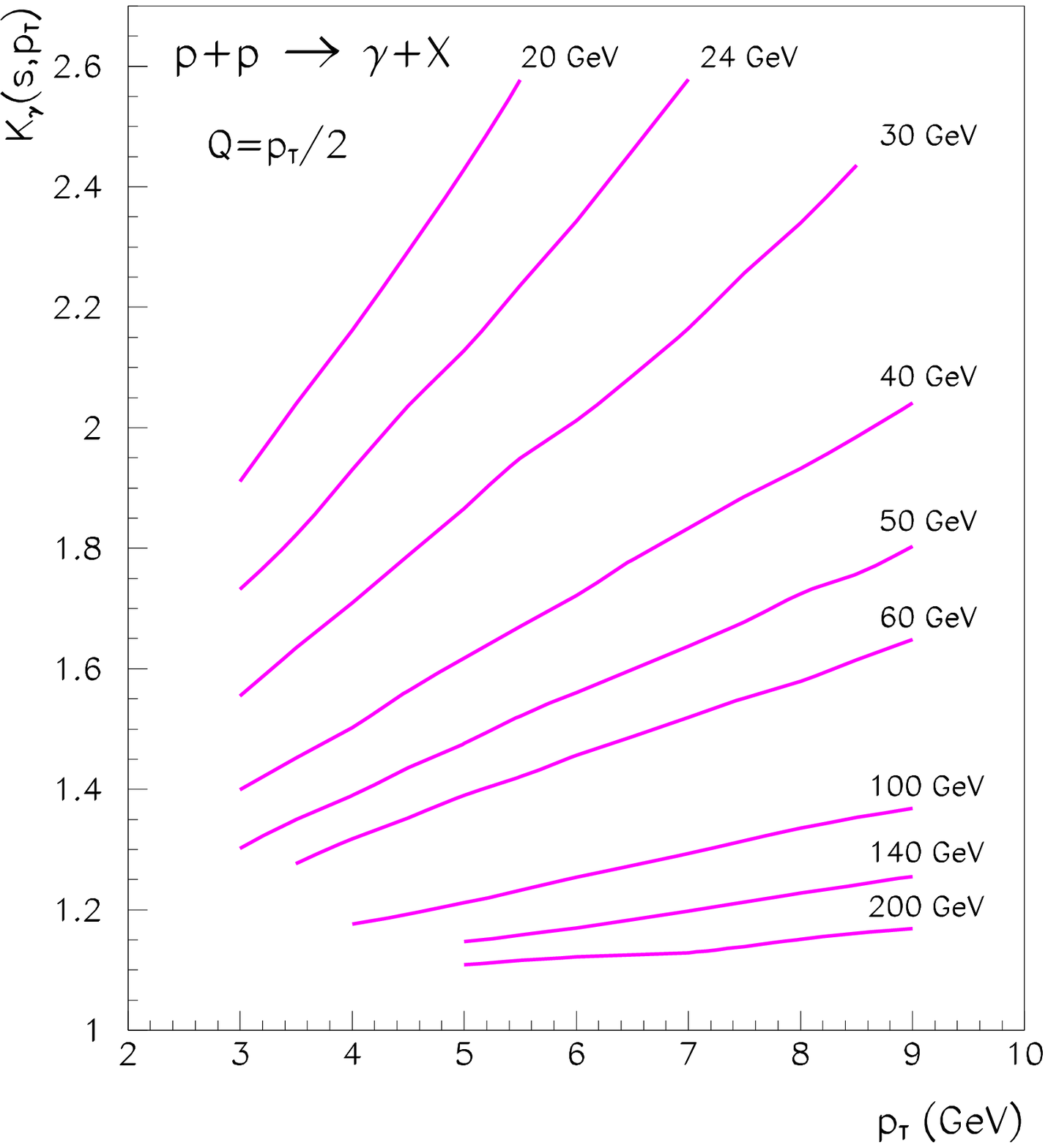}
\begin{minipage}[t]{15.cm}
{ {\bf Fig.~5.}
$K$ factor for photons,   
$K_{\gamma}(s,p_T)$ at  energies $\sqrt{s}=20 - 200$ GeV. 
}
\end{minipage}
\end{center}


\bigskip

\begin{thebibliography}{99}

\bibitem{harris}
J.W. Harris and B. M\"uller, Ann. Rev. Nucl. Part. Sci. {\bf 46} (1996) 71.

\bibitem{field89}
R.D. Field, {\it Applications of Perturbative QCD}, Frontiers in Physics
Lecture, Vol. 77 (Addison-Wesley, Reading, MA, 1989). 

\bibitem{qm99}
Proceedings of Quark Matter'99 Conference, Torino,
Nucl. Phys.  A {\bf 661} (1999) 1.

\bibitem{Aur99f}
P. Aurenche, M. Fontannaz, J.Ph. Guillet, B.A. Kniehl, E.Pilon, and M. Werlen,
Eur. Phys. J. C {\bf 9} (1999) 107.

\bibitem{Aur99p}
P. Aurenche, M. Fontannaz, J.Ph. Guillet, B.A. Kniehl, and M. Werlen,
Eur. Phys. J. C {\bf 13} (2000) 347.

\bibitem{kidonakis99}
N. Kidonakis and J.F. Owens, Phys. Rev. D {\bf 61} (2000) 094004.

\bibitem{florian99}
D. de Florian and Z. Kunszt, Phys. Lett. B {\bf 460} (1999) 184.

\bibitem{guo96}
X. Guo and J. Qiu, Phys. Rev. D {\bf 53} (1996) 6144.

\bibitem{eskola96}
K.J. Eskola, V.J. Kolhinen, and P.V. Ruuskanen,
Nucl. Phys. B {\bf 535} (1998) 351;
K.~J.~Eskola, V.J. Kolhinen, and C.A. Salgado,
Eur. Phys. J.  C {\bf 9} (1999) 61.

\bibitem{EskWang95}
K.J. Eskola and X.N. Wang, Int. J. Mod. Phys. A {\bf 10} (1995) 3071.

\bibitem{wong98}
C.Y. Wong and H. Wang, Phys. Rev. C {\bf 58} (1998) 376.

\bibitem{Soper1}
S.D. Ellis, Z. Kunszt, and D.E. Soper,
Phys. Rev. Lett. {\bf 62} (1989) 726;
Phys. Rev. D {\bf 40} (1989) 2188;
Phys. Rev. Lett. {\bf 69} (1992) 1496.

\bibitem{Soper2}
Z. Kunszt and D.E. Soper,
Phys. Rev. D {\bf 46} (1992) 192.

\bibitem{Kdrell} 
M.C. Abreu {\it et al.} Phys. Lett. B {\bf 410} (1997) 337.

\bibitem{Bass}
S.A. Bass and B. M\"uller, Phys. Lett.  B {\bf 471} (1999) 108.

\bibitem{MRST98} 
A.D. Martin, R.G. Roberts, W.J. Stirling, and R.S. Thorne,
Eur. Phys. J. C {\bf 4} (1998) 463; hep-ph/9907231.

\bibitem{pzfbl00}
G. Papp, Y. Zhang, G. Fai, G.G. Barnafoldi, and P. Levai, in preparation.

\bibitem{Aversa}
F. Aversa, P. Chiappetta, M. Greco, and J.Ph. Guillet,
Phys. Rev. Lett. {\bf 65} (1990) 401, Z. Phys. C {\bf 49} (1991) 459.

\bibitem{Ellis}
 R.K. Ellis, J.C. Sexton, Nucl. Phys. B {\bf 269} (1986) 445.
 
\bibitem{EKScode} 
http://zebu.uoregon.edu/\~{\ }soper/soper.html
 
\bibitem{EKS92}
S.D. Ellis, Z. Kunszt, and D.E. Soper, Phys. Rev. Lett. {\bf 69} (1992) 3615.

\bibitem{Huston}
J. Huston, Proc. of the 29th International Conference on
High-Energy Physics (ICHEP'98), Vancouver, Canada, 1998,
hep-ph/9901352.

\bibitem{BKK}
J. Binnewies, B.A. Kniehl, and G. Kramer, Phys. Rev. D {\bf 52} (1995) 4947.

\bibitem{Moncode}
http://home.cern.ch/\~{\ }monicaw/phonll.html

\bibitem{plf00}
G. Papp, P. L{\'e}vai, and G. Fai, Phys. Rev. C {\bf 61} (2000) 021902(R).

\end{thebibliography}
\end{document}